\documentstyle[12pt]{article}
\begin{document}
\title{ Self duality and soldering in odd dimensions}
\author{
{ R.Banerjee} \thanks{e-mail: rabin@boson.bose.res.in}  and 
{ S.Kumar} \thanks{e-mail: kumar@boson.bose.res.in} \\
S.N.Bose National Centre For Basic Sciences\\
Block JD Sector III\\
Salt Lake, Calcutta-700091  \\
India
}  
\date{}
\maketitle
\bigskip
\begin{abstract}
    Using the recently  developed soldering formalism we highlight 
certain featuers of quantum mechanical models. The complete correspondence
between these models and self dual field theoretical models in odd 
dimensions is established. The distinction between self duality 
and self dual factorisation in these dimensions is clarified.  
\end{abstract}
\newpage
 Self dual models in odd dimensions, characterised by the presence of 
Chern-Simons terms \cite{dunne}, have been in vogue for quite sometime
 \cite{TPN,DJ}. Interest in these models has been rekindled by noting
 their relevance in higher dimensional bosonisation \cite{R}. Some new
 results in this connection were reported in \cite{RW} by using the
 concept of soldering \cite{X}. Interestingly, several facets of
 self dual field theories in odd dimensions may be better appreciated 
and understood by looking at their one dimensional counterpart 
- the so called topological quantum mechanics \cite{DJT}. 
In this paper we discuss the concepts of self duality and soldering
in the context of topological quantum mechanics. Some familiar results
are explained in a different setting, leading to fresh insights. 
 This analysis is extended to the self dual field theoretic models 
in odd dimensions. Some new results are reported clarifying, 
in particular, the distinction between self duality and self dual 
factorisation.

 The quantum mechanical topological models are governed by the 
lagrangian \cite{DJT},

\begin{equation}
{\cal {L}} = \frac{m}{2}{\dot{\vec{x}}}^2 + e\dot{\vec{x}}.\vec{A}(\vec{x})
-eV(\vec{x})
\label{lag}
\end{equation}
implying the motion of a particle of mass $m$ and charge $e$ in the external
electric $(-\partial_iV)$ and magnetic $(\partial_iA_j-\partial_jA_i)$ fields.
For the simplest explicitly solvable model \cite{DJT}, the motion is two
dimensional $(i=1,2)$ and rotationally symmetric in a constant magnetic
field(B) and a quadratically scalar potential so that, 

\begin{eqnarray}
A_i &=& -\frac{1}{2}\epsilon_{ij}x_jB \nonumber \\
V   &=& \frac{k}{2}{x_i}^2  \nonumber
\end{eqnarray}
The lagrangian (\ref{lag}) therefore simplifies to (setting e=1),

\begin{equation}
 {\cal {L}} = \frac{m}{2}{\dot{x_i}}^2 + \frac{B}{2}\epsilon_{ij}x_i
\dot{x_j} - \frac{k}{2}{x_i}^2
\label{lag1}
\end{equation}
There are some interesting features of this lagrangian. If the magnetic field
is switched off (B=0), the model represents a bi-dimensional 
harmonic oscillator,

\begin{equation}
 {\cal{L}}_{H.O} = \frac{m}{2}{\dot{x_i}}^2-\frac{k}{2}{x_i}^2
\label{ho}
\end{equation}
Now consider the motion of the particle in the absence of the electric field
so that we have, 

\begin{equation}
 {\cal {L}}_+ = \frac{m}{2}{\dot{x_i}}^2 + \frac{B}{2}\epsilon_{ij}x_i\dot{x_j}
\label{l+}
\end{equation}
Let us next illustrate the connection between (\ref{l+}) and (\ref{ho}).
Together with (\ref{l+}), consider the lagrangian (\ref{l-}) with
 an independent set of coordinates $y_i$ and where the direction of 
the magnetic field is reversed,

\begin{equation}
 {\cal{L}}_- = \frac{m}{2}{\dot{y_i}}^2 - \frac{B}{2}\epsilon_{ij}y_i\dot{y_j}
\label{l-}
\end{equation}
It is now possible to combine (\ref{l+}) and (\ref{l-}) by the soldering 
formalism \cite{RW,BG}. Consider the following transformation, 

\begin{equation}
\delta x_i = \delta y_i = \eta_i
\label{varq}
\end{equation}
which effects the changes,

\begin{equation}
\delta{\cal L}_{\pm} = J_{\pm i}\dot{\eta_i}
\label{varl}
\end{equation}
where, 

\begin{equation}
J_{{\pm}{i}}(z) = m\dot{z_i} \pm B\epsilon_{ji}z_j 
\label{cur1}
\end{equation}
and $z_i = x_i ,y_i$. Introduce the soldering field $W_i$ transforming as 

\begin{equation}
\delta{W}_i = \dot{\eta_i}
\label{ax1}
\end{equation}
Then the first iterated lagrangian, 

\begin{equation}
{\cal L}^{(1)} = {\cal L}_+ + {\cal L}_- - (J_{+i}(x) + J_{-i}(y))W_i
\label{itl1}
\end{equation}
transforms as, 

\begin{equation}
\delta{\cal L}^{(1)} = -2m\dot{\eta_i}W_i
\label{varl1}
\end{equation}
Including the term ${W_i}^2$ now yields an invariant lagrangian,

\begin{equation}
{\cal L} = {\cal L}^{(1)} + m{W_i}^2  \quad\quad;\quad\quad  
 \delta{\cal L} = 0
\label{finl1}
\end{equation}
Since ${W_i}$ is an auxiliary variable, it is possible to eliminate it by
using the equation of motion, 

\begin{equation}
W_i = \frac{1}{2m}(J_{+i} + J_{-i})
\label{solax}
\end{equation}
The solution is compatible with the variations (\ref{varq}) and 
(\ref{ax1}). Inserting (\ref {solax}) into (\ref{finl1}), 
the final soldered lagrangian is obtained,

\begin{equation}
{\cal L} = \frac{m}{2}{\dot {q_i}}^2 - \frac{B^2}{2m}{q_i}^2   \quad\quad ;\quad
q_i = \frac{1}{\sqrt 2}(x_i - y_i)
\label{soldlag}
\end{equation}
which is no longer a function of $x$ and $y$ independently, but
 only on their difference. Identifying $k$ with $\frac{B^2}{m}$, 
it is found that (\ref{soldlag}) exactly maps on to (\ref{ho}). 
This exercise shows how two identical 
particles moving in the presence of magnetic fields  with same magnitudes 
but opposite directions simulate the effect of a single particle moving 
in the presence of a quadratic scalar potential.
\\
    It is easy to supplement the above lagrangian analysis by the familiar 
hamiltonian formulation \cite {DJT}. The hamiltonian corresponding to 
(\ref{l+}) is given by, 

\begin{equation}
H_+ = \frac{1}{2m}{(p_i + \frac{B}{2}\epsilon_{ij}x_j)}^2
\label{h+}
\end{equation}
where $p_i$ is the conjugate momentum,

\begin{equation}
p_i = \frac{\partial {\cal L}_+}{\partial{\dot{ x_i}}} = m{\dot{ x_i}} - \frac{B}{2}\epsilon_{ij}x_j
\label{pi}
\end{equation}
\\
Making a canonical transformation, 
\begin{equation}
p_{\pm} = p_1 \pm \frac{B}{2}x_2
\label{p+}
\end{equation}
\begin{equation}
x_{\pm} = \frac{1}{2}x_1 \mp \frac{1}{B}p_2
\label{x+}
\end{equation}
we obtain, in the new canonical variables,

\begin{equation}
H_+ = \frac{1}{2m}{p_+}^2 + \frac{1}{2}{B}^2{x_+}^2 
\label{newh}
\end{equation}

The hamiltonian is that of the usual H.O. It is however expressed only in terms of $(x_+,p_+)$ while the other canonical pair $(x_-,p_-)$ gets eliminated.
The fact that the two dimensional lagrangian (\ref{l+}) simplifies to a one
dimensional oscillator (\ref{newh}) is essentially tied to its symplectic 
structure. Likewise (\ref{l-}) yields the hamiltonian for the H.O expressed 
only in terms of the canonical set $(x_-,p_-)$. Thus the combination of 
(\ref{l+}) and (\ref{l-}) should yield a two dimensional H.O. 
which is precisely shown by the soldering mechanism leading to 
(\ref{soldlag}).

It is worthwhile to mention that the massless version of (\ref{lag1}),

\begin{equation}
{\cal L}_0 = \frac{B}{2}\epsilon_{ij}x_i\dot{x_j }
-\frac{k}{2}{x_i}^2
\label{l0}
\end{equation}
also yields a one dimensional H.O. The simplest way to realise
 this is by eliminating either $x_1$ or $x_2$ in favour of the other.
 In this sense it is similar to (\ref{l+}). Correspondingly, a 
soldering scheme can be developed.
\\[0.5cm]
Going back to the original lagrangian (\ref{lag1}), it is well known \cite{DJT}
from a hamiltonian analysis that the model corresponds to two decoupled 
one-dimensional oscillators described by the canonical pairs $(x_{\pm},p_{\pm})$and frequencies $\omega_\pm$ where,

\begin{eqnarray}
p_{\pm} & =& {\sqrt {\frac{\omega_\pm}{2m{\Omega}}}}p_1 \pm {\sqrt {\frac{{\omega_\pm}m\Omega}{2}} }x_2       \nonumber \\
x_{\pm} & =& {\sqrt {\frac{m{\Omega}}{2{\omega_\pm}} }}x_1 \mp \frac{1}{\sqrt 
{{\omega_\pm}m\Omega}}p_2    \nonumber \\
\omega_{\pm} & =& \Omega \pm \frac{B}{2m} \quad\quad   ;    \quad\quad \Omega = \sqrt { \frac{B^2}{4m^2} + \frac{k}{m}  }    
\label{R} 
\end{eqnarray}

These are the analogous of (\ref{p+}), (\ref{x+}). While the
 hamiltonian analysis reveals the decoupling of (\ref{lag1}) 
into the two one-dimensional osillators, the soldering formalism will 
explicitly demonstrate the reverse process. Let us
therefore consider the following {\it independent} lagrangians, 

\begin{equation}
{\cal L}_- = \frac{1}{2}(\omega_-\epsilon_{ij}x_i{\dot {x_j}} - {\omega
_-}^2{x_i}^2)
\label{L}
\end{equation}
\begin{equation}
{\cal L}_+ = \frac{1}{2}(-\omega_+\epsilon_{ij}y_i{\dot {y_j}} - {\omega
_+}^2{y_i}^2)
\label{M}
\end{equation}

These lagrangians are similar to the previous cases (see, for instance 
(\ref{l0})), except that the frequencies are different $\omega_{\pm}$.
 As stated before both these represent one dimensional harmonic 
oscillators but there are
 two points which ought be stressed. The equations of motion are given by, 

\begin{equation}
x_i = \frac{1}{\omega _-}\epsilon_{ij}{\dot{x_j}}
\label{A}
\end{equation}
\begin{equation}
y_i = - \frac{1}{\omega _+}\epsilon_{ij}{\dot{y_j}}
\label{B}
\end{equation}
Define a dual field as, 
$$
 {\tilde {x_i}} = \frac{1}{\omega}\epsilon_{ij}\dot {x_j}
$$
The duality property is only on-shell because, 

\begin{eqnarray}
{\tilde {\tilde{x_i}} }  = \frac{1}{\omega}\epsilon_{ij}{\tilde {\dot {x_j}} }
 = x_i   \nonumber 
\end{eqnarray}
requires the use of the equation of motion. In this sense, therefore,
 equations (\ref {A}) and (\ref{B}) characterise self and antiself 
dual solutions, respectively. Moreover, as discussed in \cite{RW, BG}, 
it is possible to interpret the lagrangians (\ref{L}) and (\ref{M}) 
as chiral oscillators with varying frequencies $\omega_\pm$ 
rotating in clockwise and anticlockwise directions. Thus the 
ubiquitous role of self duality and chriality becomes apparent in these
models. The process of soldering will combine the dual 
aspects of these symmetries to yield a new model expressed in
 terms of the composite variable 
$(x-y)$.
Under the transformations,
$$
\delta x_i = \delta y_i = \eta_i
$$
the lagrangians undergo the variations, 

\begin{eqnarray}
\delta{\cal L}_{\mp} = \epsilon_{ij}J_{\mp j }{\eta_i}\quad\quad ;
 z  = x,y  \nonumber \quad\quad ;
J_{\mp i }(z)  = \omega_\mp(\pm \dot{z_i} + \omega_\mp\epsilon_{ij}z_j)
\nonumber
\end{eqnarray}

Inserting the auxiliary variable $W_i$ transforming as, 
$$
{\delta W}_i = \epsilon_{ij}\eta_j
$$
it is posible to construct, in analogy with (\ref{finl1}), the following 
lagrangian,
$$
{\cal L} = {\cal L} _-(x) + {\cal L} _+(y) + W_i({J_i}^-(x) + {J_i}^+(y))
- \frac{1}{2}({\omega_+}^2 +{\omega_-}^2){W_i}^2
\nonumber 
$$
This expression is on shell invariant. Eliminating $W_i$ by using the equation
of motion, the above lagrangian is recast in the manifestly invariant form,
$$
{\cal L} = \frac{1}{2} {\dot{X_i}}^2 + \frac{1}{2}(\omega_+ - \omega_-)\epsilon
_{ij}X_i\dot{X_j} - \frac{1}{2}\omega_+\omega_-{X_i}^2
\nonumber
$$

\begin{equation}
X_i = {\sqrt {\frac{\omega_+\omega_-}{{\omega_+}^2 +{\omega_-}^2}}} \quad(x_i - y_i)
\label{S}
\end{equation}
where use has been made of the onshell conditions (\ref{A}, \ref{B}). 
Identifying the frequencies $\omega_\pm$ with those occurring 
in (\ref{R}) we find,

\begin{eqnarray}
\omega_+ - \omega_- &=& \frac{B}{m} \nonumber \\
 {\omega_-}{\omega_+} &=& \frac{k}{m}
\label{T}
\end{eqnarray}
After a suitable scaling it is now simple to observe that the lagrangian in 
(\ref{S}) exactly reproduces (\ref{lag1}).

The above exercise therefore shows, in a precise manner, how the self and 
anti-self dual (or, alternatively, the left and right chrial) oscillators
combine to yield the model (\ref{lag1}). For identical frequencies $(\omega_+
=\omega_- =\omega)$, the epsilon term in (\ref{S}) vanishes so that 
the lagrangian (\ref{ho}) is obtained, a result found earlier \cite{RW,BG} 
in a different context. This is also expected since (\ref{ho}) was 
derived directly from a soldering of (\ref{l+}) and(\ref{l-}), 
models which are equivalent to (\ref{L}) and 
(\ref{M}) with identical frequencies.
\\[0.2cm]
       We conclude our discussion on the topological quantum mechanics 
by pointing out that the equation of motion obtained from (\ref{S}) 
fatorises into its dual (chiral) components as follows,
$$
(\omega_+\delta_{ij} + \epsilon_{ij}\partial_t)(\omega_-\delta_{jk} - \epsilon_{jk}\partial_t)X_k = 0
\nonumber
$$
The possibility of this factorisation is ingrained in the soldering of (\ref{L})
and (\ref{M}) (with equations of motions (\ref{A}) and (\ref{B}), respectively)
to yield the final structure (\ref{S}).
\\[0.1cm]
It is now straightforward to extend the preceding analysis to odd dimensional
field theories. In this context we recall that (\ref{lag1}) had been regarded 
\cite{DJT} analogous to the lagrangian density for three dimensional topologically massive electrodynamics (Maxwell-Chern-Simons theories) in the Weyl 
$(A_0=0)$ gauge,
$$
{\cal L} = \frac{1}{2} {\dot{\vec A}}^2 + \frac{\mu}{2}{\dot{\vec A}}\times
{\vec A} - \frac{1}{2}{(\vec\nabla \times \vec A)}^2
\nonumber
$$
In our scheme of things, however, we should interpret (\ref{lag1}) to be the analouge of the topologically massive electrodynamics augmented by the usual mass
term,

\begin{equation}
{\cal L}_{S} = \frac {1}{2}{ A_\mu}{ A^\mu}- \frac {\theta}{2m^2}
\epsilon_ {\mu \nu \sigma}{\partial ^\mu}{ A^\nu}{A^\sigma}
- \frac {1}{4m^2}{ A _{\mu \nu}}{A^{\mu \nu}}
\label{V}
\end{equation}

$$
A_{\mu \nu}=
\partial_{[\mu}A_{\nu]}
$$
in the limit where all spatial derivatives are neglected \cite{dunne}. 
Correspondingly, (\ref{L}) and (\ref{M}) would be interpreted 
as the analogous of the self and anti-self dual models \cite{DJ}

\begin{eqnarray}
{\cal L}_{-}(g)= \frac {1}{2}g_{\mu}g^{\mu}  -
\frac {1}{2m_-}\epsilon _{\mu \nu \lambda} g^\mu
\partial ^\nu g^\lambda
\label{SD2d}
\end{eqnarray}

\begin{eqnarray}
{\cal L}_{+}(f) = \frac {1}{2}f_{\mu}f^{\mu} +
 \frac {1}{2m_+} \epsilon _{\mu \nu \lambda} f^\mu
\partial ^\nu f^\lambda
\label{ASD2d}
\end{eqnarray}
once again in the limit where all spatial derivatives are ignored. Since 
(\ref{L}) and (\ref{M})were soldered to yield (\ref{lag1}), it is 
natural to think that (\ref{SD2d}) and (\ref{ASD2d}) should be 
soldered to yield (\ref{V}).
This is indeed true as will now be shown. Indeed the soldering mechanism leads
to an equivalent lagrangian (\ref{V}) with the following identifications,

\begin{eqnarray}
{A_\mu} &=& { f_\mu} - {g_\mu }   \nonumber \\
 m_+ -m_-& =& \theta \nonumber \\
 m_+m_-& =& m^2
\label{nm3}
\end{eqnarray}
which is highly reminiscent of the quantum mechanical analysis.
 To begin with the soldering  consider the gauging of the following symmetry,

\begin{equation}
\delta f^{\mu }
= \delta g^{\mu}=\\
\epsilon ^{\mu \sigma \lambda}
\partial _\sigma \alpha_ \lambda
\label{fdvr}
\end{equation}
Under these transformations , the anti-self and self dual 
lagrangians change as ,

\begin{equation}
\delta {\cal L}_{\pm} ={ J_{\pm}}^{\mu \nu}{ \partial_\mu}{\alpha _\nu} 
\end{equation}
where the currents are given by,

\begin{equation}
{J}_{\pm}^{\rho \sigma}(h)
= \epsilon ^{\mu \rho \sigma}
h_{\mu} \pm 
\frac {1}{m_\pm}
\partial ^{[\rho} h^ {\sigma]}
\quad\quad ;\quad h=f,g
\label{current}
\end{equation}
Next, the soldering field $ B_{\rho \sigma} $, which is a 
two form gauge field transforming as,

\begin{equation}
\delta B_{\rho \sigma} =\partial_\rho \alpha_\sigma -
\partial_\sigma \alpha _\rho - \frac{1}{2M}(\partial_\rho
\epsilon_{\sigma\eta\zeta} - \partial_\sigma\epsilon_
{\rho\eta\zeta})\partial^\eta\alpha^\zeta
\label{axvr}
\end{equation}
is introduced. In analogy with the quantum mechanical analysis, 
it is possible to define a modified lagrangian,

\begin{equation}
{\cal L} = {\cal L}_+ + {\cal L}_- + \frac{1}{2}B^{\rho \sigma}
B_{\rho \sigma} - \frac{1}{2}B^{\rho \sigma}(J^+_{\rho \sigma}(f)
+ J^-{\rho \sigma}(g))
\label{w}
\end{equation}
which transforms as,

\begin{equation}
\delta {\cal L} =\frac {1}{2M}\epsilon^{\mu \nu \lambda}
[(f_\lambda + \frac{1}{m_+}\epsilon_ {\lambda \rho \omega}f^{\rho \omega})
 + (g_\lambda - \frac{1}{m_-}\epsilon_ {\lambda \rho \omega}g^{\rho \omega})
]\times
[\epsilon_{\mu \sigma \beta}\partial_{\nu}\partial^{\sigma}\alpha^{\beta}]
\label{s}
\end{equation}

where,

\begin{equation}
M = \frac{m_+m_-}{m_+ - m_-}
\label{mass}
\end{equation}
\\
[0.1cm]
     It is useful to observe that (\ref{s}) vanishes for $ m_+ = m_-$. 
In that case the lagrangian becomes gauge invariant under the 
transformations (\ref{fdvr}). The auxiliary $B_{\rho \sigma}$ field 
can be eliminated from (\ref{w}) in favour of the original variables 
by using the equation of motion. The final lagrangian then turns out 
to be the Proca model with the basic field as 
$A_\mu = f_\mu - g_\mu$. Incidentally, following our system of ignoring 
spatial derivatives, the Proca model just reduces to the bi-dimensional
harmonic oscillator (\ref{ho}). Likewise, (\ref{ASD2d}) and (\ref{SD2d})
with $m_+ = m_-$ can be identified with (\ref{l0}) and its dual partner.
The soldering in the latter case leads to (\ref{ho}) which provides 
another correspondence between the quantum mechanical and field theoretical 
models. In the same spirit it may be realised that (\ref{l+}) would be the
analogue of the Maxwell-Chern-Simons theory. The equivalence of (\ref{l+})
with (\ref{l0}) therefore indicates a similar connection between the 
MCS theory and the self dual model (\ref{SD2d}) - a fact which has 
been established earlier using various approaches \cite{DJ,RR}.

Coming back to the soldering mechanism for different masses $(m_+ \neq 
m_-)$, it is seen that the variation (\ref{s}) is non zero. It is possible to make further alterations to (\ref {w}) so that the new lagrangian is gauge invariant. Such alterations invariably require terms involving derivatives of
the soldering field $B_{\rho \sigma}$. In that case a simple elimination of this field in favour of the other fields , by using the equations of motion
, would not be possible. That would defeat our purpose of recasting the 
lagrangian in terms of the difference $(f_\mu - g_\mu )$;
a form in which it would be manifestly gauge invariant leading
to a new structure. It is now observed that by relaxing the
requirement of gauge invariance to be only on-shell, in which case 

\begin{equation}
f_{\mu}=-\frac {1}{m_+} \epsilon _{\mu \nu \lambda} \partial ^\nu f^\lambda
\label{m1}
\end{equation}

\begin{equation}
g_{\mu}=+\frac {1}{m_-} \epsilon _{\mu \nu \lambda} \partial ^\nu g^\lambda
\label{m}
\end{equation}
then $\delta {\cal L}$ in (\ref{s}) indeed vanishes and the lagrangian is 
gauge invariant. This is reminiscent of the quantum mechanical analysis.\\
     Returning to a description in terms of the original variables is now 
possible by eliminating $B_{\rho \sigma}$, which acts as an auxiliary 
field, from (\ref{w})\\

\begin{equation}
B_{\rho \sigma} =\frac{1}{2}(J^+_{\rho \sigma}+J^-_{\rho \sigma})  
\label{l}
\end{equation}

It should be mentioned that this solution is compatible with the variation
(\ref{axvr}) since
$$\frac{1}{2}\delta (J_+^{\rho \sigma}(f) + J_-^{\rho \sigma}(g)) =
\partial_\rho \alpha_\sigma -\partial_\sigma \alpha _\rho 
- \frac{1}{2M}(\partial_\rho\epsilon_{\sigma\eta\zeta} -
\partial_\sigma\epsilon_{\rho\eta\zeta})\partial^\eta\alpha^\zeta 
=
\delta B_{\rho \sigma}
$$
Inserting the solution (\ref{l}) in (\ref{w}) and using the on-shell 
conditions (\ref{m1}) and (\ref{m}) one obtains the Chern-Simons-Proca 
lagrangian (\ref{V}) with the identifications (\ref{nm3}).
 \\
         A straightforward extionsion of the above analysis in $d =4k-1$
dimensions would lead to the soldering of the self and anti self-dual 
lagrangians,
\begin{eqnarray}
{\cal L}_{+}&=&\frac{1}{2m_+}\frac{1}{2k!} \epsilon_
{\mu_1}\cdots{\mu_{2k-1}}{\lambda_1}\cdots{\lambda_{2k}}
f^{\mu_1\cdots\mu_{2k-1}}\partial^{[\lambda_1}f^{\lambda_2
\cdots\lambda_{2k}]}  + \nonumber \\
&+&\frac{1}{2}f_{\mu_1
\cdots\mu_{2k-1}}{f}^{\mu_1\cdots\mu_{2k-1}}
\label{diff ASDN}
\end{eqnarray}

\begin{eqnarray}
{\cal L}_{-}& =&- \frac{1}{2m_-}\frac{1}{2k!}{ \epsilon_
{\mu_1}\cdots{\mu_{2k-1}}{\lambda_1}\cdots{\lambda_{2k}}}
 g^{\mu_1\cdots\mu_{2k-1}}\partial^{[\lambda_1}g^{\lambda_2
\cdots\lambda_{2k}]}  + \nonumber \\
&+&{\frac{1}{2}}g_{\mu_1
\cdots\mu_{2k-1}}g^{\mu_1\cdots\mu_{2k-1}}
\label{diff SDN}
\end{eqnarray}

to yield the new lagrangian\\

\begin{eqnarray}
{\cal L}_{S} &=& \frac{1}{2}A^{\mu_1 \cdots \mu_{2k-1}}
A_{\mu_1 \cdots \mu_{2k-1}} -\frac{1}{2M}
\frac{1}{(2k-1)!}\epsilon ^{\mu_1 \cdots \mu_{2k-1} \sigma_1
\cdots \sigma_{2k}}A_{\mu_1 \cdots \mu_{2k-1}}
\partial _{\sigma_1}A_{ \sigma_2 \cdots \sigma_{2k}}\nonumber \\
& - &\frac{1}{2.2k}\frac{1}{m_+m_-}F_{ \sigma_1 \cdots \sigma_{2k}}
F^{ \sigma_1 \cdots \sigma_{2k}}
\label{final mscp}
\end{eqnarray}

where ,
$$
A_{\mu_1 \cdots \mu_{2k-1}} = f_{\mu_1 \cdots \mu_{2k-1}} 
- g_{\mu_1 \cdots \mu_{2k-1}}  
$$
and,
$$
 F^{\sigma_1 \cdots \sigma_{2k}}=\partial^{[\sigma_1}A^{\sigma_2
\cdots \sigma_{2k}]}
$$
where in the latter expression antisymmetrisation is done with respect
to all the indices in the square bracket and $M$ is as defined in (\ref{mass}).
Note that the basic variables $(f,g)$ are $ (2k-1)$-form fields. 
For identical masses , $m_+ = m_-$, the generalised Proca model 
is obtained .\\
   Next let us discuss the factorisability property. As noted 
in \cite{PK}\footnote{there is a sign error in this ref} the equation of motion following from (\ref{V}) factorises as,

\begin{equation}
[g^{\mu}_\sigma \mp (\frac{1}{m_\pm}){\epsilon_{\sigma}}^{\lambda
\mu}\partial_\lambda ]
[g^{\rho}_\mu \pm (\frac{1}{m_\mp}){\epsilon_{\mu}}^{\nu \rho}
\partial _\nu ] A_{\rho} =0
\label{factor}
\end{equation}

For identical masses $(m_+ = m_-)$ this reduces to the Proca equation. The 
structure of the factorisation has led to the claim that the massive modes in these models satisfy the self duality condition. That this is not so is easily shown. Consider, for simplicity, the following generating functional for 
the Proca lagrangian,

\begin{equation}
 Z_{P}[j,J] = \int DA_{\mu} e^{ -\frac{1}{2}\int
[{\cal L}_P + {\bf A}_\mu j^\mu +\tilde{\bf A}_\mu J^\mu]
{d^3}x }
\label{pfn}
\end{equation}
where, the dual has also been introduced,

\begin{equation}
{\tilde A}_\mu = \frac{1}{m} \epsilon_ {\mu \nu \lambda}
\partial^{\nu}{\bf A}^\lambda
\label{dual}
\end{equation}

The result of the Gaussian integration is ,

\begin{equation}
Z_{P}[j,J] = e^{-\frac {1}{2}\int(j_\mu + \frac{1}{m}\epsilon_ {\mu \lambda \sigma}
\partial^{\lambda}J^{\sigma})C^{\mu \nu}(j_\nu + \frac{1}{m}\epsilon_
{\nu \alpha \beta}\partial^{\alpha}J^\beta)}
\label{pfng}
\end{equation}
where,
$$
C^{\mu \nu}(x,y)=\frac{2}{(m^2 + \Box)}[g^{\mu \nu} + \frac{1}{m^2}
\partial^\mu \partial^\nu]\times \delta(x-y)
$$
It is now easy to calculate the relevant correlation functions,

\begin{equation}
< A_\eta(x) A_\xi(y)> = C_{\eta \xi}(x,y)
\label{gfn2}
\end{equation}
\begin{equation}
< A_\eta(x) A_\xi(y)> = <\tilde{A}_\eta(x)\tilde{A}_\xi(y)>
+ \frac{2}{m^2}{\bf g}_{\eta \xi}{\bf \delta} (x-y)
\label{gfn}
\end{equation}
\begin{equation}
 <{A_\eta}(x){\tilde A}_\xi(y)> = \frac{2}{m}. \frac{1}{(m^2+\Box)}
\epsilon_{\eta \sigma \xi }\partial ^\sigma \delta (x-y)
\label{gfn1}
\end{equation}

 It is seen that all the correlation functions cannot be related 
modulo only local terms. Thus it is not possible to interpret 
 $A_\mu = \tilde A_\mu $ operatorially . Hence the 
$A_\mu$ field cannot be regarded as self dual.\\

The origin of the structure of the factorisation in (\ref{factor}) is 
understood from the soldering analysis performed earlier. The two factors 
correspond to the self dual and anti self dual modes, not in the model 
(\ref{V}), but rather in the models (\ref{SD2d}) and (\ref{ASD2d}), 
respectively. It is the soldering mechanism that has precisely combined
these modes from distinct models with fields $f_\mu$ and $g_\mu$ to yield the new model (\ref{V}) with the field $A_\mu = f_\mu - g_\mu$. This new
field $A_\mu$ is altogether a seperate entity which lacks the original 
symmetry properties. 

     To conclude, we have used the soldering formalism to abstract 
different quantum mechanical models starting from the basic harmonic
oscillator. The analysis was directly extended to field theory in odd
dimensions where exploiting the self duality property, the soldering
formalism was effectively employed. It was striking that all the results
and interpretations found in the quantum mechanical examples had the 
exact analogues in the field theory.

\end{document}